\documentclass{jfm} 

\usepackage{amsmath} 
\usepackage{amssymb} 
\usepackage{array}
\usepackage{enumitem}
\usepackage{graphicx}
\usepackage{natbib} 
\usepackage{booktabs} 
\usepackage{easytable}
\usepackage{syntonly}
\usepackage{xcolor}
\usepackage{marginnote}
\usepackage{rotating}
\usepackage{lineno}
\DeclareRobustCommand{\s}{\ifmmode\mathsection\else\textsection\fi}

\renewcommand{\Re}{\ensuremath{{Re}}}

\newcommand{\Ri}{\ensuremath{{Ri}}}
\newcommand{\Rf}{\ensuremath{{Rf}}}
\renewcommand{\Pr}{\ensuremath{{Pr}}}
\newcommand{\Prt}{\ensuremath{{Pr_t}}}

\newcommand{\Fr}{\ensuremath{{Fr}}}
\renewcommand{\S}{\ensuremath{{S_\ast}}}
\newcommand{\Sh}{\ensuremath{{S_\ast}}}

\newcommand{\Reb}{\ensuremath{{Re_b}}}
\newcommand{\Res}{\ensuremath{{Re_s}}}
\newcommand{\Rs}{\ensuremath{{Re_s}}}

\renewcommand{\vec}[1]{\mathbf{#1}}

\DeclareMathSymbol{\varchi}{\mathord}{letters}{88}

\makeatother \title{
Implications of inertial subrange scaling for stably stratified mixing 
}

\author[{G. D. Portwood, S. M. de Bruyn Kops and C. P. Caulfield}]{G. D. Portwood${}^{1,2}$, S. M. de Bruyn Kops${}^1$, C. P. Caulfield ${}^{3,4}$}
\affiliation{
${}^1$ Department of Mechanical and Industrial Engineering,
       University of Massachusetts Amherst, Amherst, Massachusetts 01003, USA \\ 
${}^2$ X-Computational Physics Division, 
       Los Alamos National Laboratory, Los Alamos, New Mexico 87545, USA \\
${}^3$ BP Institute, University of Cambridge, Cambridge, 
       CB3 0EZ, UK \\
${}^4$ Department of Applied Mathematics \& Theoretical
       Physics, University of Cambridge, Cambridge, 
       CB3 0WA, UK  }

\begin{document}

\maketitle

\begin{abstract} 
We investigate the effects of turbulent dynamic range on active scalar mixing in stably stratified turbulence by adapting the theoretical passive scalar modelling arguments of \citet{beguier78} and  demonstrating their usefulness through consideration of the results of direct numerical simulations of statistically stationary homogeneous stratified and sheared turbulence.  
By analysis of inertial and inertial-convective subrange scaling, we show that the relationship between active scalar and turbulence time scales is predicted by the ratio of the Kolmogorov and Obukhov-Corrsin constants provided mean flow parameters permit the two subrange scalings to be appropriate approximations.  
We use the resulting relationship between timescales to parameterise an appropriate turbulent mixing coefficient $\Gamma \equiv \chi/\epsilon$,  defined here as the ratio of available potential energy ($E_p$) and turbulent kinetic energy ($E_k$) dissipation rates.  
With the analysis presented here, we show that $\Gamma$ can be estimated by $E_p,E_k$ and a universal constant provided an appropriate Reynolds number is sufficiently high.  This large Reynolds number regime appears here to occur at $\Reb \equiv \epsilon / \nu N^2 \gtrapprox 300$ where $\nu$ is the kinematic viscosity and $N$ is the characteristic buoyancy frequency.  We propose a model framework for irreversible diapycnal mixing with robust theoretical parametrisation and asymptotic behaviour in this high-$\Reb$ limit.  
\end{abstract}

\newpage

\section{Introduction}
Stably stratified turbulence may  be thought of as a model flow that is potentially useful
for understanding aspects of geophysical and engineering processes in 
many regions of the
oceans and atmosphere.  In particular, stratified turbulence describes the relatively small-scale
dynamics, in length and time, at which turbulence and irreversible mixing occurs.
Interpreting and modeling such idealised flows is a primary avenue for the
development and calibration of global circulation and basin-scale models, for
which mixing plays a leading-order role 
in global energy budgets
\citep[e.g.][]{ferrari09,jayne09,gregg18b}.  
However, even in idealised stably stratified
flow configurations, parameterised modelling of mixing has emerged as a
challenge due to the potential for dependence on a wide range of non-independent
parameters \citep{ivey18,gregg18b,caulfield21}.

Parallel to the growing recognition of the importance of modelling such
small-scale processes in stratified turbulence for broader geophysical applications, stratified turbulence has been
increasingly observed to exhibit certain quantitatively similar small-scale dynamics to isotropic
turbulence in some parameter regimes specifically when the buoyancy Reynolds number,
$\Reb$, is large enough \citep{gargett81,lindborg06a,debk15}. 
Of course, this does not mean that stratified turbulence is equivalent in all respects, not least because stratification inevitably introduces anisotropy.
We
consider this parameter in more detail in the next section, 
but in simple terms it quantifies the
dynamic range of turbulent length scales neither directly affected by large-scale
stratification nor by molecular viscosity.  Indeed, it is widely acknowledged that the details of the
large scale, or outer, mean scale, dynamics do not strongly affect the small scale
dynamics provided there is sufficient scale separation between such large and
small scales.   While small scale dynamics are certainly not independent of large scales
\citep{corrsin58,durbin91},  the assumption that sufficient scale separation
induces statistically-independent small-scales has proven to be a valuable tool
in the modelling of anisotropic turbulent flows because it allows for the
application of theoretical models based on statistical symmetries, e.g.,
isotropy and homogeneity, to dynamic modelling.  

Just to take one example,
Kolmogorov-Obukhov scaling depends on the assumptions of local isotropy and
homogeneity \citep{kolmogorov41,oboukhov41a} yet it has been observed to be a
good approximation for anisotropic flows provided the scale separation is
sufficiently large between anisotropic turbulence scales and the dissipative
viscous-diffusive scales \citep{champagne70,gargett84,saddoughi94,shen02}. 
However,
not
all stably stratified flows exhibit sufficient scale separation
for
the foregoing assumption to be made \citep{jackson14,debk15}.   

This phenomenology is useful because it motivates the adaptation of existing
models for isotropic turbulence to stably stratified turbulence.  For instance, by analysis of
Obukhov-Corrsin and Kolmogorov scaling, \citet{beguier78} observed that the
passive scalar timescale $\tau_\phi$ and turbulence timescale $\tau$ couple at
sufficiently high Reynolds number such that they can be related by universal
constants, i.e.
\begin{equation}
\frac{\tau_\phi}{\tau} \equiv \frac{E_\phi \epsilon}{\epsilon_\phi E_k} = \frac{\beta}{C},
\label{eq:beg_rel_passive}
\end{equation}
where $E_\phi$ is the turbulent scalar variance, $E_k$ is the turbulent kinetic
energy, $\epsilon_\phi$ and $\epsilon$ are their respective dissipation rates and
$\beta,C$ are the Obukhov-Corrsin and Kolmogorov constants, respectively.  The
assumption that the time scale ratio is a constant is commonly applied to model
scalar dissipation in Reynolds-averaged Navier-Stokes (RANS) models
\citep[e.g.][]{newman81,ristorcelli06}. 

In modeling mixing in stratified flows, the applicability of such a universal relation is
interesting due to  its implications for modeling the irreversible diapycnal
flux, $j_b$.  In such models, equilibrium assumptions are typically prescribed
such that the irreversible diapycnal flux is equal to a particular definition of the available potential
energy dissipation rate \citep{osborn72,peltier03,caulfield21} %\cpc{(essentially a an appropriately scaled scalar variance dissipation rate)} 
such that coupling of turbulent and scalar dynamics, as suggested by \citep{beguier78}, can be an
insightful relation for modeling scalar dynamics. 

The importance of accurately modelling such irreversible scalar fluxes in the ocean
has motivated significant research of various idealised flows using an array of modelling frameworks.  Perhaps
most notable, estimating the irreversible flux from the scalar gradient
via a turbulent diffusivity, $\kappa_b$, is the \textit{de facto} standard
approach \citep[e.g.][]{ivey91,barry01,maffioli16,salehipour15}, with
\begin{equation}
j_b = \kappa_b N^2
\end{equation}
where $N$ is is an appropriately defined large scale buoyancy frequency.  A common parametrisation of $\kappa_b$
has been suggested by \citet{osborn80}, which asserts that $\kappa_b$ be
related to the irreversible rate of dissipation of kinetic energy 
via a turbulent flux coefficient $\Gamma_b$ such that 
\begin{equation}
  \kappa_b = \Gamma_b
\epsilon/N^2 \rightarrow \Gamma_b \equiv \frac{j_b}{\epsilon} ,
\label{eq:gammadef}
\end{equation}  By analysis of energy equations for a simple stratified flow
model at stationary conditions, \citet{osborn80} suggested that $\Gamma_b$ be  a
constant $\leq 0.2$. More recently, the prescription of parameter dependence
for $\Gamma_b$ has emerged as a necessary improvement to turbulent diffusivity
models due to increasing evidence that a constant coefficient for $\Gamma_b$ appears not to be at all suitable.
Parametrisations of $\Gamma_b$ in this framework have emerged to
be difficult due to potential dependence on multiple parameters,
which may well themselves be correlated \citep{caulfield21},
which difficulties lead to often contradictory subclosures \citep{monismith18}.  
While alternative models exists, such as
mixing-length models \citep{odier09,ivey18} and flux transport models, they
have not as yet gained widespread usage in large-scale simulations.
%Turbulent
%diffusivity models additionally allow the instantaneous irreversible flux to be
%calculated from instantaneous turbulence quantities, as is frequently necessary
%oceanographic measurements.  

Therefore, as the first principal objective of this research, we consider formally adapting the \citet{beguier78} relationship
\eqref{eq:beg_rel_passive} to stably stratified turbulence  in order to
model irreversible diapycnal mixing dynamically. To evaluate this hypothesis,
we consider simulations of \emph{stationary homogeneous stratified and sheared
turbulence} (S-HSST).  S-HSST is an idealised model flow configuration that enables easy
adjustment of the range of length scales (the dynamic range) available for a
locally-isotropic subrange to form without changing other dimensionless flow
parameters such as characteristic Froude or Richardson numbers, as defined precisely in the next section. 
Such a fundamental flow configuration is well suited to the evaluation of the 
%to evaluating the 
effects of locally-isotropic scaling theories \citep{shih00,chung12}.
%In evaluating 
To test
the hypothesis that the \citet{beguier78} relation 
%may 
can 
be applied to stably stratified turbulence, we perform numerical experiments of S-HSST in parameter space extremes previously inaccessible to computation.  Therefore, a second principal objective of this work is to report on the characteristics of S-HSST at very large Reynolds numbers and time scales, quantitatively described in subsequent sections.

To achieve these objectives, the rest of the paper is organised as follows.
In \s \ref{sec:background},
we discuss
a length-scale based framework for stratified and sheared
turbulence and use it to derive a mixing model hypothesis based on
the arguments of \citet{beguier78}. 
In 
\s \ref{sec:simulation}, we then describe numerical simulation experiments of
%we provide some background discussion on 
{\emph{homogeneous stratified and sheared
turbulence}} (HSST).
%which is used to design direct numerical simulation experiments as discussed . 
In a Reynolds number parameter-space, 
we then present ensemble-averaged dynamics 
%are  observed 
in
\s \ref{sec:results}, 
%evaluation of the 
evaluate the mixing model 
%is performed 
in \s \ref{sec:scaling},
and finally draw our
conclusions in \s \ref{sec:conclusions}.

\section{Theoretical background}
\label{sec:background}
\subsection{Parametric framework}
\label{sec:framework}
Homogeneous sheared and stratified turbulence, HSST,  is assumed to satisfy the
Navier-Stokes equations subject to the non-hydrostatic Boussinesq
approximation.  The dimensional equations for the fluctuations relative to the
planar means are 
\begin{subequations}
  \label{eq:nsb_fluct_set}
\begin{align}
    \frac{\partial \vec{u}}{\partial t} +
    (\vec{u} \cdot \nabla)\vec{u} 
    & = -
    \frac{1}{\rho_0} \nabla p -
    z \frac{\partial \vec{u}}{\partial x} \frac{d\bar{u}_x}{dz} - 
    u_z \frac{d\bar{u}_x}{dz} \hat{\vec{x}}-
    \frac{g_z}{\rho_0}\rho\hat{\vec{z}} +
    \nu \nabla \cdot \nabla \vec{u},
    \label{eq:fluct_vel_eq}\\
    \frac{\partial \rho}{\partial t} + 
    (\vec{u} \cdot \nabla) \rho 
    &= -
     z \frac{\partial \vec{\rho}}{\partial x} \frac{d\bar{u}_x}{dz} 
    - u_z\frac{d\bar{\rho}}{dz} +
    D \nabla \cdot \nabla \rho , 
    \label{eq:fluct_scalar_eq}\\
    \nabla \cdot \vec{u} &= 0,
    \label{eq:cont_eq}
\end{align}
\end{subequations}
where $\vec{u} = (u_x, u_y, u_z)$ is the velocity vector in the coordinate
system $(x,y,z)$ and functional dependencies have been dropped for convenience,
and $\rho$ is the fluctuating density field so that the total density is
\begin{equation}
\rho_t = \rho_0 + \overline{\rho}(z) + \rho(x,y,z,t),
\end{equation}
with $\overline{\rho}(z) = z d\overline{\rho}/dz$ being the ambient density
that is constant in time and has a uniform gradient antiparallel to
gravitational acceleration $g_z$.  Similarly, $\bar{u}_z= z d\overline{u}_x/dz=z S$
is the mean velocity and $S\equiv d\bar{u}_x/dz$ is the mean shear. The pressure decomposition is analogous to that of
density and velocity so that $p$ is the fluctuating mechanical pressure
relative to the temporally-constant planar mean.  Internal energy and scalar
transport affecting density have been combined into a single evolution equation
for $\rho$ with $D$ the molecular diffusivity, $\hat{\vec{x}}$ 
and $\hat{\vec{z}}$ are  the unit
vectors in the $x$- and $z$-directions respectively, and $\nu$ is the kinematic viscosity.

\subsubsection{Parametrisation}
\label{sec:param}
Dimensional analysis suggests that \eqref{eq:nsb_fluct_set} can be described in
terms of at least four nondimensional parameters.  
As
we wish to interpret the dynamics in terms of the dynamic range available for various aspects of the flow, 
we
consider parametrisation in terms of the ratios of length scales.

Here we use the convention of defining an approximate outer scale of turbulence, the large-eddy length scale, and the (viscous) Kolmogorov length scale as
\begin{equation}
  L_{LE} \equiv E_k^{3/2}/\epsilon \text{ and } L_K \equiv (\nu^3/\epsilon)^{1/4} ,
  \label{eq:scales_1}
\end{equation}
where $E_k \equiv \langle \vec{u} \cdot \vec{u} \rangle/2$, $\epsilon \equiv
\nu \langle \nabla \vec{u} : \nabla \vec{u} \rangle$ , the operator `$ : $' denotes the double inner product and the notation $\langle
\cdot \rangle$ indicates an ensemble average.  

The (buoyancy) Ozmidov length scale
\begin{equation}
  L_O\equiv(\epsilon / N^3)^{1/2} 
  \label{eq:scales_2}
\end{equation}
defines the lower limit of length scales significantly affected by buoyancy forces,
where $N$ is the (background) buoyancy frequency, defined here as
\begin{equation}
    N^2 \equiv -\frac{g_z}{\rho_0} \frac{d\overline{\rho}}{dz} . \label{eq:n2}
\end{equation}
With an explicit shear scale applied in HSST, the Corrsin length scale \citep{corrsin58} is thought to characterise the lower limit of scales affected
by shear, and is defined as
\begin{equation}
  L_C \equiv (\epsilon / S^3)^{1/2}.
  \label{eq:scales_3}
\end{equation}
%and scales 
Scales smaller than $L_C$ have been suggested to define a locally isotropic regime of length scales in the absence of additional smaller-scale affects of the mean-flow \citep{saddoughi94}.

Given the scales defined in (\ref{eq:scales_1}), (\ref{eq:scales_2}) and (\ref{eq:scales_3}), we choose to describe turbulence by the following three parameters 
\begin{equation}
  \Rs \equiv \frac{\epsilon}{\nu S^2}= \left(
  \frac{L_C}{L_K}\right)^{4/3}
  \text{, }  
  \Ri \equiv \frac{N^2}{S^2} =\left (
  \frac{L_C}{L_O}\right )^{4/3}\text{ and } 
  \Fr \equiv \frac{\epsilon}{N E_k}= \left (
  \frac{L_O}{L_{LE}}\right )^{2/3},\label{eq:res}
\end{equation}
in a fluid with a fixed Prandtl number $Pr\equiv \nu/D$
as the required fourth parameter (see \citet{mater14} for more details on parameterisations).
%The 
We choose the
shear Reynolds number, $Re_s$, 
%is chosen 
since it describes the range of length scales associated with isotropy for stationary flows with Richardson number, $Ri \leq 1$ \citep{mater14}. 
Alternative Reynolds numbers can also be useful to consider for comparison to other flow regimes, where the turbulent Reynolds number and buoyancy Reynolds numbers are defined, respectively as
\begin{equation}
  \Re \equiv \frac{E_k^2}{\nu \epsilon}
 =\left (
  \frac{L_{LE}}{L_K}\right )^{4/3}
  ,  \ \Reb \equiv \frac{\epsilon}{\nu N^2} 
  =\left (
  \frac{L_O}{L_K}\right )^{4/3} .
\end{equation}

In flows driven by mean shear, the energetic stationarity of the flow is thought to be characterised by 
%the gradient Richardson number 
$Ri$
\citep{jacobitz97}, measuring the
scale separation between the driving shear scales and stabilising buoyancy
scales.  It is thought that \textit{turbulent} flow configurations persist
where $Ri<1$, where we distinguish between sustained turbulent flows and results derived from or interpreted in terms of linear
hydrodynamic instability theory \citep[see][for more discussion]{zhou17}.
Indeed, stratified turbulent simulations with prescribed shear scales have
revealed that the characteristic gradient Richardson number associated with stationarity has
been observed to take a Reynolds number insensitive value of
\begin{equation}
 \Ri \approx 0.15 - 0.2 \text{,}
  \label{eq:ri_stat}
\end{equation}
as observed in stratified homogeneous shear flow \citep{shih00,holt92,portwood19}. 

Furthermore, in shear-dominated flows, energy production due to shear naturally induces a strong coupling between shear and outer length scales such that the shear parameter $\S$, here defined 
in terms of $E_k$ (as opposed to in terms of $q\equiv\langle \vec{u} \cdot \vec{u} \rangle$) as
\begin{equation}
  \S \equiv \frac{E_k S}{\epsilon}=\left (
  \frac{L_{LE}}{L_C}\right )^{2/3},
\end{equation}
tends to a constant between 5 and 6 at sufficiently high Reynolds numbers \citep{jacobitz97,shih00} 

However the foregoing turbulent parameters are likely 
insufficient
to characterise active scalar dynamics, a specific point we wish to investigate in this paper. The smallest scales of the scalar are thought to be characterised by the Batchelor length
scale $L_B$, 
\begin{equation*}
  L_B \equiv (\nu D^2 / \epsilon)^{1/4},
\end{equation*}
which coincides with $L_K$, at unity Prandtl number as considered in this work.

Obukhov-Corrsin similarity, which is either applicable at high Froude
number or at scales smaller than those substantially affected by buoyancy, i.e. for scales smaller than
$L_O$, implies an outer-scale scalar length scale
\begin{equation}
  L_{OC} \equiv   E_p^{3/2} \epsilon^{1/2} / \chi^{3/2},\label{eq:loc}
\end{equation}
where $E_p$ is the available potential energy, $\chi$ is its irreversible
dissipation rate, defined in the linearly stratified limit to be 
\begin{equation}
  E_p \equiv \left \langle \frac{g^2/\rho_0^2}{2N^2}\rho^2 \right \rangle \text{ and }
\chi \equiv D \left \langle \frac{g^2/\rho_0^2}{N^2} | \nabla \rho | ^2 \right \rangle,
\label{eq:ep_chi_def}
\end{equation}
respectively, demonstrating that $\chi$ in this context may also be interpreted 
as the scaled destruction rate of buoyancy variance \citep{caulfield21}.   Therefore, the range of anisotropic scales of the scalar are characterised by
\begin{equation}
N_\ast \equiv \left( \frac{L_{OC}}{L_O} \right)^{2/3}= \frac{N E_p}{\chi}. 
\end{equation}
\citet{portwood19} observed this parameter to approach unity in the high $Re_b$ limit for flows where $Pr=1$, and \citet{mater14} suggests $N_\ast$  more descriptively parameterises turbulent mixing.
\subsection{A mixing model constructed from analysis of subrange scaling}
\label{subsec:hyp}
Local isotropy is a state wherein statistical symmetries of multi-point
 statistics, such as homogeneity, isotropy and stationarity, are present in a
spatio-temporal localised region \citep{monin75}.  In the presence of
anisotropic integral scales, the conditions of local isotropy are defined, in
the spatial sense, by a \textit{quasi-equilibrium} range of scales %which 
that are
sufficiently smaller than integral scales \citep{kolmogorov41}.  

This is a prerequisite of classical inertial subrange
scaling arguments \citep{kolmogorov41,oboukhov41a,corrsin51}. These state that
scales within the \textit{quasi-equilibrium} regime exist wherein turbulence
is independent of the effects of viscosity. In shear flows, conditions for the
application of local isotropy are thought to be valid at scales smaller than
$L_C$ \citep{corrsin58,uberoi57}.  Kolmogorov scaling
\citep{kolmogorov41,kolmogorov62} in the presence of anisotropic outer-scales
has been revealed to coincide at scales consistent with local isotropy
\citep{champagne70,saddoughi94}.   The equivalent local isotropy justification
of the  Kolmgorov subrange may be applied to Obukhov-Corrsin passive scalar
scaling \citep{oboukhov49,corrsin51} such that it is fit to describe an active
scalar below scales affected by stratification \citep[][pg 391]{monin75}.
Therefore, the spectral density of potential energy is expected to be
determined by the energy dissipation rates and a wavenumber, i.e.
\begin{equation}
  \hat{E}_{p}^{OC}(|\vec{k}|) = \beta \chi \epsilon^{-1/3} |\vec{k}|^{-5/3} f_L^{OC}(|\vec{k}| L_{C})f_v^{OC}(|\vec{k}| L_B),
\label{eq:oc_def3}
\end{equation}
where $\vec{k}$ is the wavenumber vector, $\beta$ is the Obukhov-Corrsin
constant, the functions denoted by $f^{OC}_v$, $f^{OC}_L$ are the viscous and
outer scale correction functions that are necessary without %assumptions of
applying assumptions of local isotropy.  Furthermore, according to \cite{oboukhov41a}, the kinetic
energy spectra
\begin{equation}
  \hat{E}_{k}^{K}(|\vec{k}|) = C \epsilon^{2/3} |\vec{k}|^{-5/3} f_L^K(|\vec{k}| L_{C})f_v^K(|\vec{k}| L_K), 
\label{eq:k_def}
\end{equation}
where $C$ is the Kolmogorov constant, and $f^{K}_L$ and $f^{K}_v$ are the
correction functions.  Following the analysis of \citet{beguier78}, by integration of
\eqref{eq:oc_def3} and \eqref{eq:k_def} when $L_C 
\gg |\vec{k}|^{-1} \gg L_K, L_B$,
it is possible to obtain
\begin{equation}
\Pi \equiv \frac{E_p \epsilon}{E_k \chi}= \frac{\beta}{C} .
\label{eq:beg}
\end{equation}
Therefore, the explicit relation for modeling the irreversible flux from
(\ref{eq:beg}) is
\begin{equation}
j_b \approx \chi \approx \frac{E_p \epsilon}{E_k \Pi}, 
  \label{eq:hyp_rel}
\end{equation}
with the limiting behaviour, at high Reynolds number, that $\Pi \rightarrow \beta/C$. A critical simplifying assumption in the development of \eqref{eq:hyp_rel} is that $\Fr \geq O(1)$, though the imposition of more complex scaling relations which are thought to develop when $\Fr \ll O(1)$ may alternatively be imposed.
Indeed, more generally, 
\begin{equation}
\Pi = \left ( \frac{L_{OC}}{L_{LE}} \right )^{2/3} = \frac{N_\ast}{\S \sqrt{\Ri}} = N_\ast \Fr \quad .
  \label{eq:gen_hyp}
\end{equation}
%Parametrisations of $\Pi$ at moderate Reynolds numbers are investigated in  \s 6.
We stress that the assumed model spectra \eqref{eq:oc_def3} and \eqref{eq:k_def} are approximations which are difficult to observe precisely in carefully controlled experiments or simulations. Whereas these two-point scaling models are observed in flows with anistropic large scales, they are not claimed generally applicable for flows in arbitrary parameter regimes.  However the sensitivity of $\Pi$ with respect to measured deviations from the model spectra is unclear \text{a priori}, a point we discuss in \s \ref{sec:results}.

The relation \eqref{eq:hyp_rel} has profound implications for the calibration of a number of 
turbulent stratified mixing models. 
Generally, in gradient diffusion models,
\begin{equation}
    \kappa_b \approx \frac{E_p }{\Pi N} \Fr, 
\end{equation}
while gradient diffusion subject to the model proposed by \citet{osborn80} leads to
\begin{equation}
    \Gamma_b \approx \Gamma \equiv \frac{\chi}{\epsilon} =  \frac{E_p}{E_k \Pi} \text{ ,} \label{eq:gammachidef1}
\end{equation}
%Using
Combining the turbulent viscosity model of \citet{crawford82} with the turbulent diffusivity model of \citet{osborn80}, the turbulent Prandtl number may then be expressed as
\begin{equation}
\Prt = \frac{1+\Gamma_b}{\Gamma_b} \Ri = \frac{E_k \Pi+ E_p} {E_p}\Ri \text{ .}\label{eq:prtdef}
\end{equation}
Alternatively, the relevance of the parameter $\Pi$ to the turbulent Prandtl number model of \citet{venayagamoorthy10} implies that their key mixing parameter $\gamma$ may be expressed as 
\begin{equation}
    \gamma = \frac{\Pi}{2}\quad ,
\end{equation}
where $\gamma$ too has been empirically observed to assume relatively insensitive values at moderate Reynolds numbers in HSST \citep{venayagamoorthy06}.
Finally, if the objective is to construct a mixing length scale for applications in Prandtl mixing length models, such as the framework of \citet{odier09}, where $\kappa_b = l_m^2 S$, \eqref{eq:hyp_rel} may be equivalently expressed as 
\begin{equation}
l_m =  \frac{E_p }{\Pi N^2} \S \text{ .}
\end{equation}

\section{Direct numerical simulations}
\label{sec:simulation}
\label{sec:sim}
%\subsection{A review of stationary homogeneous sheared stratified turbulence}
In order to 
test the underlying hypotheses
%verify the hypothesis 
described in \s \ref{subsec:hyp} leading to the irreversible mixing model embodied in (\ref{eq:hyp_rel}), we must consider a stratified model flow wherein the behaviour of $\Pi$ may be 
%observed 
evaluated as a 
function of dynamic range associated with isotropy.  The consideration of S-HSST is justified by the emergence of a Reynolds number as the only independent non-dimensional descriptive flow parameter due to two distinct phenomenologies: (i) the coupling of the outer turbulence scales to that of the shear production  mechanism such that $\S \approx 5$ and (ii) the gradient Richardson number assuming a balanced apparently critical value that maintains a stationary balance of energy production and dissipation mechanisms. Both phenomenologies were discovered in DNS for this flow configuration by the investigations of Shih et al. \citep[][]{shih00,shih05} and subsequently verified more recently in a broader Reynolds number space in simulations where the Richardson number is controlled to maintain constant turbulent kinetic energy \citep{portwood19}. We therefore utilise the flows presented in \citet{portwood19} to verify the modeling framework presented in \s \ref{subsec:hyp} and relevant 
associated consequences highlighted
later in this manuscript.

\subsection{Numerical method}
The homogeneous stratified shear system defined by \eqref{eq:fluct_vel_eq} and \eqref{eq:fluct_scalar_eq} is solved in a triply periodic domain with a Fourier pseudo-spectral method. Timestepping is performed with a fractional step method where non-linear terms are advanced with a third-order Adams-Bashforth scheme.  Aliasing is treated by a sharp-spectral filter at $15/16 k_{max}$, which proved to be sufficient in a-posteriori analysis.  Temporal integration of the inhomogeneous shear term is handled by the integrating factor approach of \citet[][]{sekimoto16}.  The rest of the solver is derived from the method used in \citet{hebert06b}, with further details discussed therein.

\begin{sidewaysfigure}
\vspace{38em}
\centering
  \includegraphics[]{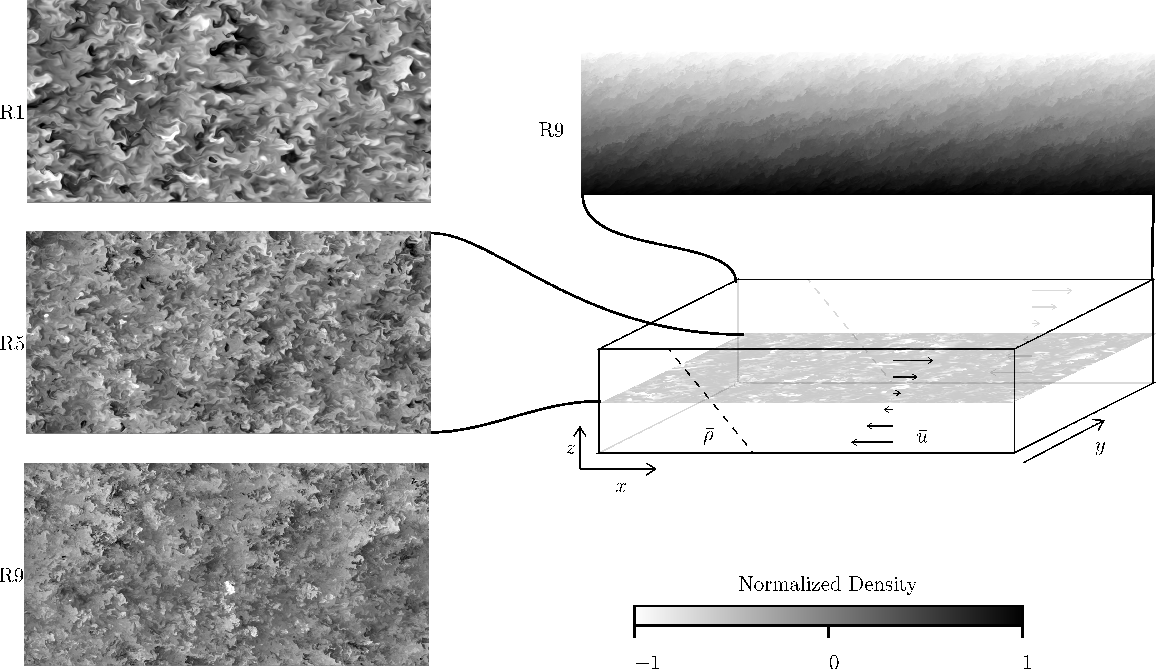}
\vspace{2em}
  \caption{Illustration of simulation domain and contour plots of density.  The three leftmost panels are slices of total density on an x-y slice, normalised by three times the variance of the fluctuating density in case R1; panels are shown for cases R1, R5, and R9. Note the that fluctuations decrease with Reynolds number while small-scale structure increases. Upper-right panel indicates total density normalised by minimum and maximum values for case R9, illustrating inclined large-scale structure \citep[c.f.][]{chung12,jacobitz16}. 
  %\cpc{\bf CPC: Reference now done to this in the revision.}
  }
  \label{fig:schem}
\end{sidewaysfigure}

The suite of simulations are designed to resolve sufficiently both the smallest and largest spatiotemporal turbulent flow scales. 
The streamwise extent, $L_x$, is is chosen to be 40 times larger than the large-eddy length scale. 
Because of the anisotropy of integral length scales that was observed in a limited parameter space study, the
cross-stream domain length $L_y$ and the vertical domain length $L_z$ are chosen such that $L_x/L_y=2$ and $L_x/L_z=4$.
A visualisation of the computational domain is featured in figure \ref{fig:schem}. 
The smallest spatial scales are
%were
resolved by enforcing  $k_{max}L_{K} \approx 2$,  where $k_\text{max}$ indicates the
maximum wavenumber supported by the domain discretisation when fully-resolved.

Finally, we have adopted a mass-spring-damper-like system (as similarly adopted in
\citet{overholt98,rao11}), to tune the Richardson number to its stationary
value by setting a constant kinetic energy target $E_t$. We note that this is similar
to the constant power criteria as used in stationary HSST by \citet{chung12}
but that here we include damping. 
The system is controlled by time-dependent variation of $Ri$ being required to satisfy the equation
\begin{equation}                                                
  c_0 S \Ri'(t)+2\alpha \omega \tilde{E}_k'(t) + \omega^2 (\tilde{E}_k(t)-1)=0, \label{eq:controls}    
\end{equation}                                              
where the prime notation denotes a temporal derivative, $\tilde{E}_k(t)\equiv
E_k/E_t$ is the normalised turbulent kinetic energy, $\omega$ is
the characteristic
frequency of oscillation, $\alpha$ is a dimensionless damping factor, and $c_0$ is a dimensionless parameter.  This
control system has been derived by assuming that the
kinetic energy follows a second order linear system \citep[e.g.][]{rao11},
and then by applying a first-order approximation to the temporal evolution of kinetic energy \citep[c.f.][]{jacobitz97} 
\begin{equation}
\tilde{E}_k'(t)\approx c_0
S (\Ri(t)-\Ri_c) \hbox{\ such that \ } \tilde{E}_k''(t) \approx c_0 S \Ri'(t). \label{eq:approxri}  
\end{equation}
The choice for the
parameter $c_0\approx -1$ is supported by \citet{jacobitz97}, the
characteristic frequency $\omega$ is determined by the large-eddy time scale which itself is proportional to $1/S$ \citep{shih00}, and finally a
damping coefficient $\alpha=1.5$ was found to work well.

\subsection{Summary of experiments}
The stationarity constraint
and the emergent empirical observation that $S_\ast \approx 5$ %determine 
lead to the natural consideration of sets of simulations
which can test the effects of variation of the Reynolds number, essentially independently of the other parameters.
The simulations presented here are largely equivalent
to those reported in \citet{portwood19}, though some cases have been run for
longer times to ensure convergence of statistics.
The various parameters of the simulations are summarised in Table
\ref{tbl:simtable}, and  some flow visualisations of the density field are shown in figure \ref{fig:schem}. Two key aspects are apparent, as noted in the caption. First, in the horizontal plane visualisations, as the Reynolds number increases, the magnitude of the density fluctuations tends to decrease, while, unsurprisingly the dynamic range of scales increases, with  enhanced smaller scale structures. Second, in the vertical plane visualisation, inclined large-scale structures are apparent, consistently with the results of \cite{chung12} and \cite{jacobitz16}.
\begin{table}
  \begin{center}
    \hspace{2em}
    \begin{tabular}{r c c c c c}
          &  \Fr   &  \Ri   &  $\Reb$  &  $\Res$  &  $N_x$  \\
\midrule
SHSST-R1  &  0.46  &  0.164  &  36     &  6      &  1024   \\
R2        &  0.46  &  0.164  &  46     &  8      &  1280   \\
R3        &  0.48  &  0.163  &  60     &  10      &  1536   \\
R4        &  0.50  &  0.158  &  80     &  13      &  1792   \\
R5        &  0.52  &  0.155  &  110    &  16     &  2048   \\
R6        &  0.48  &  0.157  &  160    &  25     &  3072   \\
    \end{tabular}
    \hfill
    \begin{tabular}{r c c c c c}
         &  \Fr   &  \Ri   &  $\Reb$  &    $\Res$  &     $N_x$  \\
\midrule
SHSST-R7  &  0.48  &  0.156  &  240  &      38  &       4096  \\
R8        &  0.47  &  0.145 &  380  &      56  &       6144  \\
R9        &  0.42  &  0.143  &  540  &      77  &       8192  \\
R10       &  0.38  &  0.150  &  760  &      115  &       9600  \\
\\
\\
    \end{tabular}
    \hspace{2em}
    \caption{Simulation parameters.  Other parameters may be inferred from definitions in the previous section.  $N_x$ is streamwise grid points in the equispaced domain discretisation.  All cases have $\Pr \equiv \nu/D = 1$.} 
    \label{tbl:simtable}
  \end{center}
\end{table}
\begin{figure}
  \begin{center}
  \includegraphics{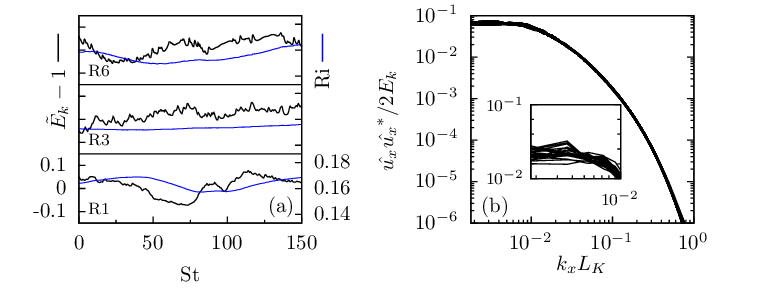}
    \caption{(a) Sample time histories of the relative kinetic energy
    fluctuation about the target energy and the Richardson number.  Smaller
    fluctuations with increasing Reynolds number were observed, but generally,
    the kinetic energy remains within 15\% of its target.  (b) Longitudinal
    streamwise velocity spectra for case R5 sampled throughout the run period
    illustrating small spectral fluctuations, where the most significant
    fluctuations are about the smallest wavenumbers.}
  \label{fig:convergence}
\end{center}
\end{figure}
Time series are shown in figure \ref{fig:convergence} to illustrate
convergence.  Richardson numbers tend to fluctuate within about 10\% of their means,
at a confidence interval at least as precise as critical Richardson numbers
reported by \citet{jacobitz97} and \citet{shih00}.  The fluctuations in the Richardson
number occur at large integral timescales such that effects of $Ri'(t)$ are
small.  Energetic fluctuations are more substantial than higher dimensional
forcing schemes \citep[i.e.][]{overholt98,rao11} that dynamically control the
flow with more granularity by accessing a broader range of turbulent length scales
compared to a single mean-scale parameter as is done here.  Nonetheless, energy
remains approximately stationary 
over the large time scales 
of the simulations.

Recalling that the critical Richardson number is an emergent quantity from the control system defined by (\ref{eq:controls}), and is not determined {\it a priori}, 
an increase of the critical Richardson number is not
observed in these simulations as suggested by \citet{holt92,shih00},
although the emergent critical value of the
Richardson numbers are broadly consistent with those reported by \citep{shih00}. Specifically, we  
observe the critical gradient Richardson number which induces statistical stationarity
in these flows to be approximately 0.16.  Similarly, we observe $\Fr\approx
0.5$ 
in all simulations, 
which implies $\Sh \approx 5$, as observed in other
studies \citep{shih00,jacobitz97}.  

Curiously, the lowest Reynolds number to maintain stationarity robustly according to our
stringent criteria corresponds approximately to case R1.  Case R1 corresponds
to $\Reb \approx 36$, which is consistent with critical values observed and
estimated for sustained three-dimensional turbulence previously
\citep{gibson80,shih05,portwood16}.

\begin{figure}
  \begin{center}
  \includegraphics{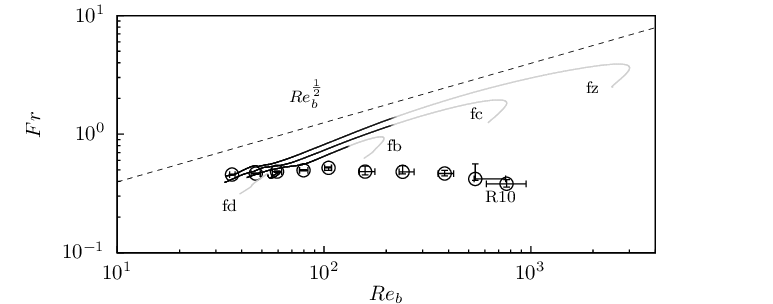}
    \caption{
    Flows in a $Fr-Re_b$ parameter space. Points indication solutions obtained in the present work. For reference, the curves indicate trajectories in parameter space of solutions for cases \textit{fz}, \textit{fc}, \textit{fb} and \textit{fd} presented in \citet{shih00} and \citet{shih05}. Darker segments of those lines indicate the reporting period $St>2$ used in that research.
    }
  \label{fig:pspace}
\end{center}
\end{figure}

For further comparison with the flows considered by 
%Shih et al. 
\cite{shih00} and \cite{shih05}, 
we plot their relevant cases against the solutions presented for this research in a $Fr-Re_b$ parameter space as shown in figure \ref{fig:pspace}. We remark that the configurations reported by
Shih and co-authors in those publications feature transitional and, crucially, variable $Fr$ when $Re_b$ is $O(100)$. 
As is apparent in the figure, $Fr$ and $Re_b$ appear to be coupled, with approximate scaling relationship $Fr \propto Re_b^{1/2}$, as shown with a dashed line. Therefore, the parameterisation of mixing by $Re_b$ cannot be made independent of $Fr$ with the data presented in \citet{shih00,shih05}. 
 We note that this explanation does not rely on any  argument that the simulations are inadequately resolved, either at small or large scales. Our perspective is qualitatively different from the perspective presented in  \citet{kunze12,gregg18b}, which suggests the $Re_b$-sensitive regime with $Re_b>100$ is due to inadequate small-scale spatial resolution in the simulations. However, it is entirely plausible that the solutions analysed in Shih et. al. suffer from inadequate large scale resolution in the domain as suggested by other reports \citep[see ][]{kunze11,kunze12,gregg18b}.  
 
Therefore, disentangling the behaviour of their simulations as $Re_b$ varies
from the effects of variation in $Fr$ is challenging, particularly when $Re_b$ (or $Re_s$) is large. It is important to remember that it was precisely for $Re_b \gtrsim 100$ that \cite{shih05} argued that the turbulent flux coefficient $\Gamma \propto Re_b^{-1/2}$, whereas it is entirely plausible that variable $Fr$ is playing a significant role. This figure highlights the novelty of our simulations in this particular flow configuration, which allow the isolated study of the effects of variations in Reynolds number from the influence of other non-dimensional turbulent parameters, and in particular at constant $Fr$.

\section{Emergent phenomena}
\label{sec:results}
\subsection{Energetics}
Just as non-dimensional parameters 
are emergent quantities in our simulations, so too
are the partitionings of potential energy
and the various components of kinetic energy.
Such partitionings are significant, not least because the ratio of potential energy to kinetic
energy, 
\begin{equation}
  R_{PK} \equiv E_p / E_k \; \text{ ,}
\end{equation}
is a critical component to mixing models wherein Reynolds number, or $\Reb$,
dependence is often omitted and the mixing is assumed to be a function of $\Ri$
\citep{osborn72,schumann95,pouquet18}.  Furthermore, in `strongly' stratified turbulent
flow, \citet{billant01} suggests %page 1649 
that there should be approximate equipartition between potential and kinetic
energy, i.e.  $R_{PK}\approx1$, an assumption also used by \citet{lindborg06a}.
It is important to remember that these
models do not account for Reynolds number effects.
Therefore,
parameterisation of such energetic partitioning is an important \textit{a priori}
validation necessary for model evaluation.

\begin{figure}
  \begin{center}
  \includegraphics{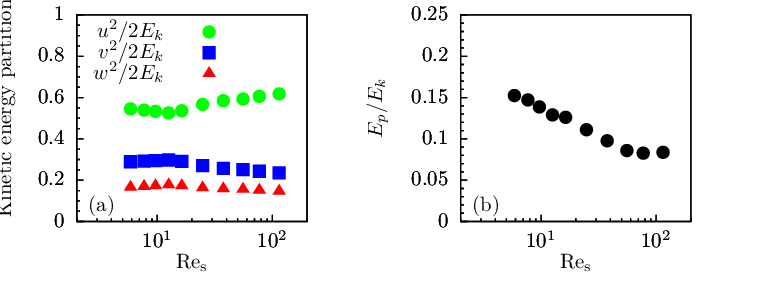}
  \caption{(a) The partitioning of kinetic energy into each velocity component
  where a value of $1/3$ would be expected for a statistically isotropic
  velocity vector. (b) The ratio of potential energy to kinetic energy, approaching asymptotic values near {0.08} for $\Res \gtrapprox 50$.  }
  \label{fig:energy}
\end{center}
\end{figure}

Recalling that the kinetic energy is the same for all cases, by construction from our definition of stationarity,  we observe a significant decrease in potential energy with
increasing dynamic range as shown in figure \ref{fig:energy}b.  In comparison,
the results of \citet{brethouwer07}, which span up to $\Reb \sim O(10)$ at much
lower Froude number in homogeneous stratified turbulence, report $R_{PK}
\approx 0.15$ at $\Reb = \Ri \Res \approx 16$, noting an asymptotic trend to this value from
$R_{PK}\approx 0.05$ at $\Reb\approx 0.1$.  Though we observe similar values at
$\Res\approx 5$ (corresponding to $\Reb \approx 30$), higher values of $\Reb$
reveal a subsequent decline in $R_{PK}$, as suggested by \citet{remmler12}.
For Reynolds numbers larger than $\Res \approx 40$ ($\Reb \approx 300$), $R_{PK}$ assumes an apparently
asymptotic value of $0.08$ in these simulations.  This relative decrease of potential
energy as a function of $\Res$ can be observed in the domain slice visualisations of
$\rho$ featured in figure \ref{fig:schem}, as there is a clear reduction in the variance of the density fluctuations.
In terms of the important question concerning the modeling of mixing, and in particular parameterising the turbulent flux coefficient $\Gamma$, \eqref{eq:gammachidef1} shows that if $R_{PK}$ approaches an asymptotic value for large $Re_s$, $\Gamma$ can exhibit parameter dependence only
if the key parameter $\Pi$ (defined in (\ref{eq:beg})) does. 

Furthermore, a distinguishing characteristic of sheared stratified turbulence, relative to
other stratified flows, is the absence of axisymmetry normal to gravity
with the result that anisotropy must be parameterised in each dimension.
Inertially-relevant anisotropy can be characterised by the variance of
individual velocity vector components as they contribute to the mean turbulent
kinetic energy.  These velocity variances resulting from anisotropic
dynamics are shown in figure \ref{fig:energy}a.
We observe turbulent kinetic energy is dominated by streamwise velocity
fluctuations with the smallest contributions coming from vertical velocity
fluctuations.  Notably distinct from axisymmetric flows, the horizontal
components of velocity variance represent approximately 80 percent of the total
contributions to kinetic energy.  The cross-stream velocity variance accounts
for only 30 percent of the total energy. The vertical variance, subject to
exchanges to potential energy, is typically suggested to vary as a function of
Froude number \citep{brethouwer07}. Here, the vertical variance accounts for approximately 20
percent of the total energy and we observe that it remains largely constant as a function of $\Reb$.

Perhaps surprisingly, anisotropy of velocity fluctuations increases with
increasing dynamic range.  That is, the streamwise velocity component becomes
increasingly dominant with increasing Reynolds number, apparently at the expense
of the cross-stream velocity variance.  One possible interpretation is that
reducing viscous effects at inertial scales as the dynamic range increases allows the flow to evolve towards an
inertial equilibrium state.  

\subsection{Dynamics}
It might be assumed that the significant energetic transitions 
at sufficiently high Reynolds number 
would be explained by accompanying clear transitions in dynamics.  Here we analyse the stationary behaviour of the dynamics relevant to the kinetic and
potential energy balances.  
The ensemble-averaged energy equations, as derived from \eqref{eq:nsb_fluct_set},  obey
\begin{align}
  \left \langle \frac{\partial {E_p}}{\partial t} \right \rangle &= B - \chi \; \text{ and } \label{eq:fluct_pe} \\
  \left \langle \frac{\partial {E_k}}{\partial t} \right  \rangle &=  P - B - \epsilon \; \text{ ,}\label{eq:fluct_ke}
\end{align}
where the turbulent production from mean shear $P=-\langle u_x u_z \,
d\bar{u}_x/{dz}\rangle$, the `buoyancy' flux  $B=\langle ({g}/{\rho_0})
{u_z} \rho \rangle$ 
have anisotropic effects on the evolution of the streamwise and vertical components of the kinetic energy $\langle u_x^2\rangle$ and $\langle u_z^2\rangle$ respectively.

Note that the left hand sides of \eqref{eq:fluct_ke}, \eqref{eq:fluct_pe}
reduce to approximately zero due to the stationarity constraint.
Therefore, the
dynamics should, at least in principle,  be describable by the turbulent flux coefficient $\Gamma$ defined in terms of $\chi$ in \eqref{eq:gammachidef1},  and a flux Richardson number, defined as
\begin{equation}
  \Rf \equiv B/P \simeq \frac{\chi}{\chi + \epsilon} = \frac{\Gamma}{1+\Gamma},  
  %\ \Gamma \text{ ,}
  \label{eq:dyn_param}
\end{equation}
since, if the flow actually enters an emergent stationary state the flux $B  \simeq \chi$,
and remembering the underlying original definition \eqref{eq:gammadef}  $\Gamma_b \simeq \Gamma$.
Furthermore,  in this circumstance, the turbulent viscosity $\nu_b$ can be naturally defined as 
\begin{equation}
    \nu_b \equiv \frac{P}{S^2} \rightarrow \Prt \equiv \frac{\nu_b}{\kappa_b} \simeq \frac{\Ri}{\Rf} \simeq \left (\frac{1+\Gamma}{\Gamma} \right ) \Ri, \label{eq:prtdef2}
\end{equation}
and so we can obtain an expression consistent with \eqref{eq:prtdef}.
Henceforth, we continue to describe the evolution of energetics by $\Gamma$ defined as in \eqref{eq:gammachidef1} in recognition of  the longstanding
application of different definitions of the turbulent flux coefficient to estimating turbulent diffusivity and turbulent viscosity
\citep{osborn72,crawford82}.
\cite{gregg18b} discuss in detail the  subtleties and potential for confusion concerning different definitions for turbulent flux coefficients, and so it must always be remembered that here we concentrate on using the definition in \eqref{eq:gammachidef1}, i.e. $\Gamma\equiv \chi/\epsilon$.

\begin{figure}
  \begin{center}
  \includegraphics{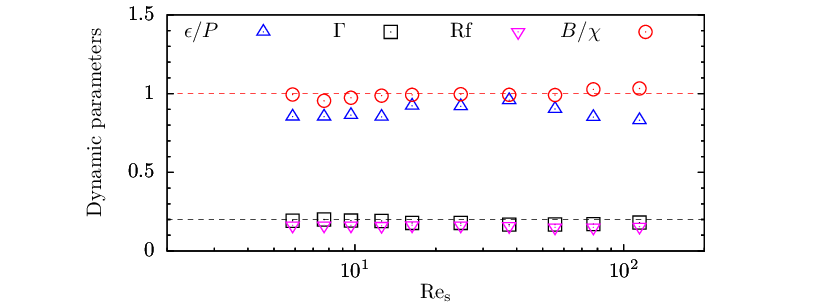}
    \caption{Relationships between various components of the potential and turbulent kinetic energy evolution equations as a function of Reynolds
    number.  The dashed line at unity corresponds to the unimposed, but emergent,
    condition that $\chi \approx B$. The dashed line at 0.2 indicates the 
    upper bound
    for $\Gamma$ postulated  by \citet{osborn80}.  Note that all dynamic ratios
    appear approximately constant as a function of Reynolds number. }
  \label{fig:dyn}
\end{center}
\end{figure}

The ratio of terms in the energetic balance is shown in figure \ref{fig:dyn}.
We observe a slight decline of $\Gamma$ from $0.2$ to approximately $0.175$, as
noted in \citet{portwood19}, until $\Res \approx 40$.
However, $\Gamma$ remains approximately
constant as $Re_s$ increases further, and in particular
we do not observe the proposed $\Reb^{-1/2}$
scaling reported in a similar flow by \citet{shih05} for their energetic regime
of $\Reb > 100$.  Remembering the data presented in figure \ref{fig:pspace}, this is perhaps unsurprising due to the apparent correlated
variation of $Fr$ and $Re_b$ in those simulations.
Due to the condition of stationarity being imposed on the kinetic energy in our simulations,
accompanying induced stationarity of the potential energy is also guaranteed,
though still plotted for reference in figure \ref{fig:dyn} 
demonstrating that $\chi \approx
B$. It is also apparent that $\Rf \equiv B/P$ is 
less than $\Gamma\equiv \chi/\epsilon$,
as expected.

\subsection{Length scales}
\begin{figure}
  \begin{center}
  \includegraphics{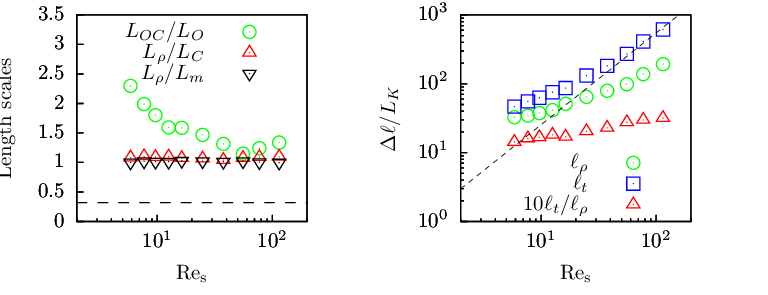}
  \caption{(a) 
  The variation with $Re_s$ of the ratios of: the Obukhov-Corrsin scalar outer-scale $L_{OC}$, defined in (\ref{eq:loc}), to the Ozmidov length $L_O$, defined in (\ref{eq:scales_2});
  the scalar mixing length $L_\rho$, defined in (\ref{eq:mixlength}),
  to the Corrsin length scale $L_C$, defined in (\ref{eq:scales_3}); and the scalar mixing length $L_\rho$ to the momentum mixing length $L_m$, also defined in (\ref{eq:mixlength}).  
  (b) The variation with $Re_s$ of: the scaling of total dynamic range, using $\Delta \ell_t$ and $\Delta \ell_\rho$, as defined
 in \eqref{eq:scalest} and \eqref{eq:scalesr} respectively; and (ten times) their ratio. 
 The expected $\Res^{4/3}$ scaling is plotted with a dashed line, 
 making apparent the anomalous scaling 
 in the range of
  potentially isotropic scales associated with the scalar.
  }
  \label{fig:scalar_out}
\end{center}
\end{figure}

Since we have observed $\Fr$ and $\Ri$ to be 
%nearly 
essentially
independent of the Reynolds number, the relatively decreasing potential energy as  $\Res$ increases for $Re_s \lessapprox 50$ implies that scales characteristic of the scalar should also exhibit a similar asymptotic trend.
The scalar outer-scale
associated with Obukhov-Corrsin similarity, $L_{OC}$ is defined in (\ref{eq:loc}). In figure
\ref{fig:scalar_out}a, we 
%show
plot the ratio of this outer-scale to the Ozmidov
length scale $L_O$, as defined in (\ref{eq:scales_2}).
As $Re_s$ increases, this ratio decreases for
$Re_s \lessapprox 40.$ For larger $Re_s$, there is 
some evidence that the ratio remains constant near unity,
suggesting of course that $L_{OC}$ anomalously scales with $L_O$.
This implies that the scaling $\chi \approx E_p N$
is valid in this high Reynolds number regime and that outer-scales 
of the scalar adjust to the regime associated by local isotropy,
which
is indeed a condition of a Obkhov-Corrsin subrange in anisotropic flows.

%refer to equation in manuscript
Length scales associated with the mixing length models suggested by \citet{odier09} can also be shown to remain constant 
in our large $Re_s$ regime, and indeed even for smaller $Re_s$.
The scalar mixing length, $L_\rho$, and 
the momentum mixing length $L_m$, both defined by \citet{odier09} as
\begin{equation}
L_m \equiv \left ( \frac{P}{S^3} \right )^{1/2}, 
\
L_\rho \equiv   \left( \frac{B}{N^2 S} \right)^{1/2} 
\; \text{ ,}\label{eq:mixlength}
\end{equation}
are coupled via the turbulent Prandtl number $\Prt$, which here is very close to one, since
\begin{equation}
    L_\rho
= L_m \left ( \frac{B S^2}{P N^2} \right )^{1/2}
= L_m \left ( \frac{\Rf}{Ri} \right )^{1/2}
\simeq \Prt^{-1/2} L_m
\; \text{ .}\label{eq:odier}
\end{equation}
Furthermore, they coincide
with the Corrsin length scale by the stationarity constraint from
\eqref{eq:fluct_ke}.
The expected scalings $L_\rho \approx L_C$ and $L_\rho \approx L_m$ are both clearly apparent in 
figure \ref{fig:scalar_out}a across a wide range of Reynolds numbers. 

The decrease of $L_{OC}$ implies that the total dynamic-range associated
with the scalar anomalously scales with the shear Reynolds number $\Res$.   We
define the total range of dynamic scales for turbulence and the scalar as 
\begin{align}
  \Delta \ell_t  &= (L_{L} - L_K) \label{eq:scalest} \; \text{  and}\\
  \Delta \ell_\rho  &= (L_{OC} - L_B) \label{eq:scalesr} \; \text{ ,}
\end{align}
respectively, and show them as a function of Reynolds number in figure
\ref{fig:scalar_out}b.  With respect to the turbulence scales, after an initial small $\Res$
transient, the range of total dynamic scales appears to increase with
$\Res^{4/3}$, as predicted by our definitions in \s \ref{sec:param}.  
%With
For the scalar, on the other hand, there is clearly a flatter-than-predicted scaling regime at high
Reynolds number which is characteristic of the transient outer scales observed
in figure  \ref{fig:scalar_out}a.  Furthermore, the range of dynamics-scales associated
with the scalar is significantly smaller than that of the turbulence,
suggesting that if a dynamic-range threshold exists, it will not occur
simultaneously for the scalar and the turbulence.

\section{Implications of inertial subrange scaling}
\label{sec:scaling}
%pg 351 monin&yaglom for the term "inertial subrange"
%top of pg 341 through top of 343 for "locally isotropic turbulence" where they deal with the shear case
%saddoughi94 gives a very similar description in his introduction
Before turning to model verification, an intermediate phenomenological test of
the universality hypothesis outlined in \s 2.3 may be performed by analysis of
the one-dimensional correlations.  The one-dimensional spectra of the potential
energy is expected to be determined by the energetic dissipation rates and a
wavenumber, i.e.  for the cross-stream spectra 
\begin{equation}
  E_{p}^{OC}(k_y) = \beta_1 \chi^{1} \epsilon^{-1/3} k_y^{-5/3} f_L^{OC}(k_y L^O_{yy})f_v^{OC}(k_y L_B) .
\label{eq:oc_def}
\end{equation}
In this expression,  $\beta_1$ is the one-dimensional Obukhov-Corrsin constant. Furthermore the functions
denoted by $f^{OC}_v$, $f^{OC}_L$ are the viscous and outer scale correction
functions which are necessary 
%without 
if no
implicit assumptions of local isotropy
are made. 
The notation $L^O_{ij}$ indicates the outer scale of locally isotropic
turbulence associated with the velocity component $i$ with respect to the
direction $j$. This is ostensibly the Corrsin length scale but is inevitably
anisotropic as discussed in
\citet{kaimal73} for the specific case of the stably stratified boundary layer.  

Equivalently,
according to \cite{kolmogorov41} for an isotropic inertial subregime, the
one-dimensional cross-stream longitudinal and streamwise transverse energy
spectra are
\begin{align}
  E_{y}^{K}(k_y) &= C_1 \epsilon^{2/3} k_y^{-5/3} f_L^K(k_y L^O_{yy})f_v^K(k_y L_K)
\label{eq:k1_def} \\
  E_{x}^{K}(k_y) &= C_1' \epsilon^{2/3} k_y^{-5/3} f_L^K(k_y L^O_{xy})f_v^K(k_y L_K)
\label{eq:k1t_def}
\end{align}
where $C_1$ and $C_1'$ are universal constants, and $f^{K}_L$ and $f^{K}_v$ are the
correction functions. The ratio of \eqref{eq:oc_def} and \eqref{eq:k1_def}
yields
\begin{equation}
  \frac{E_{\rho}^{OC}(k_y)}{E_{y}^{K}(k_y)} = \frac{\beta_1}{C_1} \frac{\chi}{\epsilon} \frac{f_L^{OC}}{f_L^K} \frac{f_v^{OC}}{f_v^K} \text{ .}
\label{eq:rat_def1}
\end{equation}

\begin{figure}
  \begin{center}
  \includegraphics{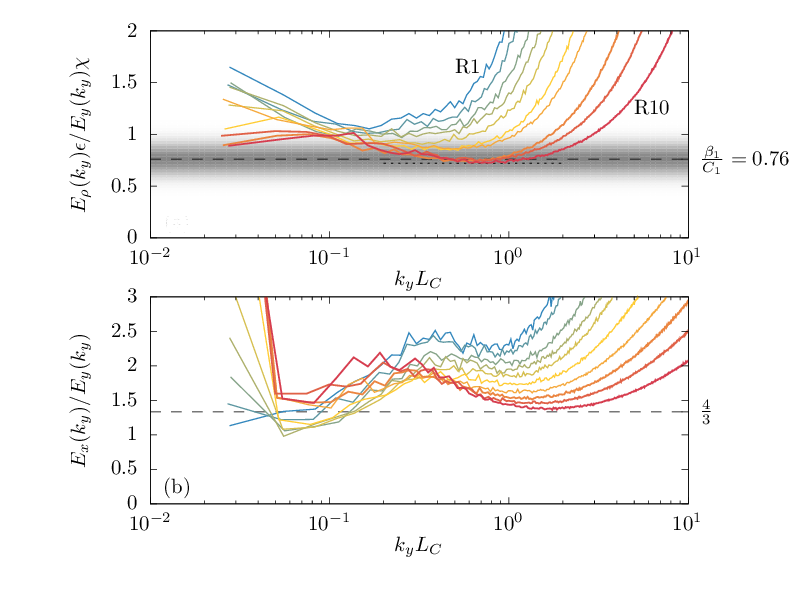}
  \caption{ (a) The compensated ratio of one-dimensional energy spectra in the
cross-stream direction.  At high wavenumbers, the spectra lie in order of
Reynolds number with R1 on the left and R10 on the right as labelled.  The
coexistence of Kolmogorov and Obukhov-Corrsin scaling, even when subject to
non-trivial correction functions, suggests a subregime below anisotropic scales
which features a plateau corresponding to the ratio $\beta_1/C_1$ of the Obukhov-Corrsin
constant to the Kolmogorov constant as predicted by \eqref{eq:rat_def} which is
expected to occur at $L_Ck_y\approx1$.  The dashed line at 0.76 indicates
measurements made in the stratified boundary layer by \citep{wyngaard71b} where
the surrounding shaded region represents its uncertainty from reported standard
deviations and the dotted line indicates an estimate of the asymptotic value of
$\beta_1/C_1$ at 0.72. (b) The ratio of one-dimensional streamwise-transverse
and cross-stream-longitudinal spectra, which have a predicted ratio given in 
\eqref{eq:43_def} at high $\Res$, expected to occur at wavenumbers a factor of
$\Ri^{-3/4}\approx 4$ greater than the locally isotropic regime shown in panel (a).
}
  \label{fig:rat_spec}
\end{center}
\end{figure}

From empirical studies of turbulence, even in the absence of the effects of
stratification and shear, we expect the correction functions $f_L$, $f_v$ to be
non-trivial \citep[e.g.][]{muschinski15}.  However, it is at least plausible that a
subregime exists wherein
\begin{equation}
  \frac{f_L^{OC}}{f_L^K} \approx 1 \; \text{  and  } \;
\frac{f_v^{OC}}{f_v^K} \approx 1 ,
  \label{eq:out_func}
\end{equation}
either due to sufficiently high scale separation 
%amongst
of isotropic scales, i.e.
$\Res \gg 1$, or due to shared functional forms such that $f_L^{OC}=f_L^K$.
Subject to either condition, it would be expected that \eqref{eq:rat_def1} 
would reduce to
%can
%be observed as
\begin{equation}
\frac{E_{\rho}^{OC}(k_y)}{E_{y}^{K}(k_y)} = \frac{\beta_1}{C_1} \frac{\chi}{\epsilon}  \; \text{.}
\label{eq:rat_def}
\end{equation}
Indeed, similar arguments would motivate the $4/3$ relation between transverse
and longitudinal velocity spectra in the presence of anisotropic outer-scales
such that $F_L^K(k_yL^O_{xy})\neq F_L^K(k_yL^O_{yy})$ in order to obtain
\begin{equation}                                                                     
  \frac{E_{x}^{K}(k_y)}{E_{y}^{K}(k_y)} = \frac{C_1'}{C_1}=\frac{4}{3} \; \text{.}    
\label{eq:43_def}                                                                    
\end{equation}  
The validity of the relation \eqref{eq:rat_def} is 
tested
with 
appropriately scaled one-dimensional spectra in figure \ref{fig:rat_spec}a, 
where we would expect the relation to
%coincide with 
hold in the locally isotropic 
wavenumber regime where $L_C k_y \sim O(1)$ \citep{saddoughi94}.
We observe a tendency for the local minima of the spectra to decrease with
increasing Reynolds number.  The minima for cases R8, R9 and R10 are all close to, and bounded below by
0.72, within the range of values reported for the
universal constants \citep{sreenivasan95,sreenivasan96}.  Furthermore, in
the stratified
boundary layer, \citet{wyngaard71b} report measurements of each constant
independently, in a flow configuration very similar to the one studied here,
which imply $0.76 \pm 0.11$,
also in agreement
with the results observed here.

The wavenumbers associated with this regime also tend toward 
%the 
smaller scales
as $\Res$ increases.  In cases R8, R9 and R10, the lower-bound for this regime
appears to remain fixed at approximately $k_yL_C=0.4$ as the right bound begins
to increase such that the \textit{flat} region of the spectra broadens.  We
note that this apparently asymptotic behaviour in cases R8, R9 and R10 occurs
while $\Res$ increases by a factor of two.
The ratio of transverse to longitudinal one-dimensional spectra, as expressed in
\eqref{eq:43_def}, is shown in figure \ref{fig:rat_spec}b.  We observe a
similar downward trend of the locally isotropic wavenumber regime with $\Res$, leading eventually to
good agreement with the $4/3$ law for case R10.

Similarly to the data shown in figure
\ref{fig:rat_spec}a, the locally isotropic regime at higher wavenumbers
becomes more consistent with the Corrsin length scale at approximately
$k_yL_C\approx2$.  The misalignment of the wavenumber regimes consistent with
\eqref{eq:rat_def} might be seen to be a consequence of the anisotropy of
outer scales, as accounted for in our definitions of anisotropic length scales
in $f_L^K$ and also in the observation that the outer scale of the scalar 
largely coincides
with the Ozmidov length. 
Therefore, it is expected that $L_{OC}/L_C \sim \Ri^{-3/4}$ and hence the wavenumber regime
associated with the four-thirds relation is expected to be approximately a
factor of four  larger than \eqref{eq:rat_def}, essentially as observed here.

\begin{figure}
  \begin{center}
  \includegraphics{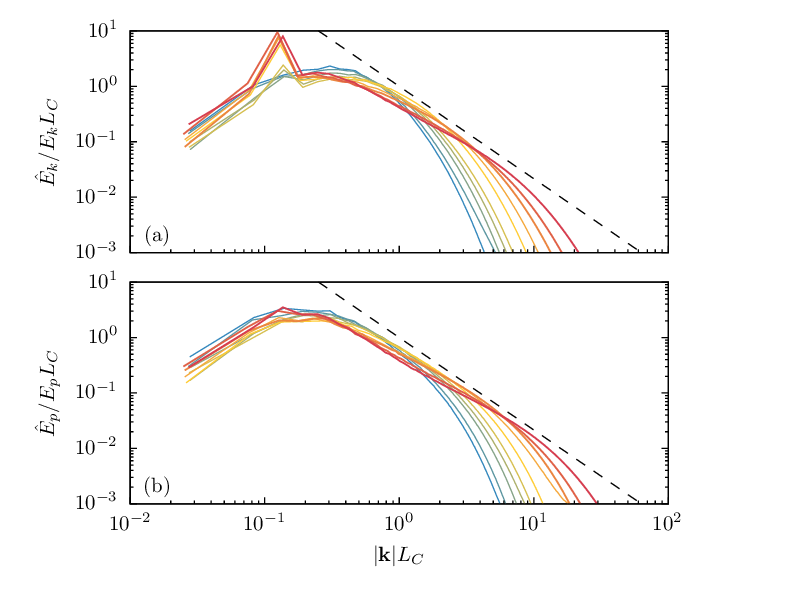}
  \caption{ 
  Three-dimensional kinetic and potential energy spectra shown in panels (a) and (b), respectively.
  The dashed black line indications a $-5/3$ slope, as assumed by the model spectra \eqref{eq:k_def} and \eqref{eq:oc_def3}.}   
  \label{fig:3dver}
\end{center}
\end{figure}

We compare measured three dimensional potential and kinetic energy spectra to the model spectra, \ref{eq:k_def} and \ref{eq:oc_def3}, in figures \ref{fig:3dver}a,b. In both panels, a $|\mathbf{k}|^{-5/3}$ curve is plotted for reference. Our intention is to verify that the proposed spectra are approximated by our analytic model spectra at high $Re_s$, in the sense that we do not observe substantial systematic deviations which would otherwise indicate alternative two-point parameterisations would be more appropriate. Here we observe a trend wherein the range of scales associated with three dimensional turbulence, $|\mathbf{k}|L_C>1$, are approximated by the $-5/3$ power law as $\Res$ increases.  Whereas deviations from the power-law behaviour are observed at all $\Res$, the quantitative sensitivity of the parameter $\Pi$ on deviations from the model spectra is unclear at this stage. Therefore, we next turn to evaluating the impact of Reynolds number on our key parameter $\Pi$.

\subsection{Model verification}
\label{sec:discuss}
In the previous section, we have 
presented evidence
that the ratio of the potential energy spectra to a
longitudinal energy spectra appears to be consistent with the underlying scaling
arguments central to coexistent Kolmogorov and Obukhov-Corrsin regimes.  The value of the key parameter $\Pi$ can be expressed in terms
of the one-dimensional universal constants, at high Reynolds number, by 
\begin{equation}
  \Pi = \frac{\beta}{C} = \frac{6}{11} \frac{\beta_1}{C_1} .
  \label{eq:pi_est}
\end{equation}
For these flows, the parameter $\Pi$ should be assumed to be universal to the
extent that $C$ and $\beta$ are.  For reasons outlined in \cite{sreenivasan91}
and also observed here, measuring individual constants from a sheared flow is
fundamentally problematic and furthermore measuring $\beta$ requires an accurate
measurement of $\chi$, which is inherently difficult in experimental flows.  

\begin{figure}
  \begin{center}
  \includegraphics{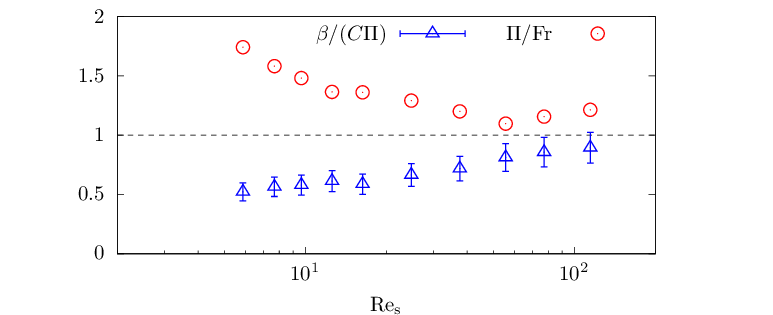}
  \caption{Verifications of relations \eqref{eq:pi_est} and \eqref{eq:pifr} for sufficiently large $\Res$. Error bars
  correspond to standard deviations associated with estimates of $C$ and
  $\beta$ from \citet{wyngaard71b}. Recall that $\Pi/Fr=N_\ast$ from (\ref{eq:gen_hyp}).}
  \label{fig:disc}
\end{center}
\end{figure}

Nonetheless,
estimates of the (3D) Kolmogorov and Obukhov-Corrsin constants in the literature
would imply that $\Pi \approx 0.42$
\citep{wyngaard71b,sreenivasan91,sreenivasan96,yeung02,debk15}
 with estimates of the individual constants  typically being
reported within approximately 15\% of each other in the literature cited.  Robust estimates
of $\Pi$ from the relation \eqref{eq:pi_est} and empirical observations in
figure \ref{fig:rat_spec} indicate $  \Pi \approx 0.39 $ which is certainly
within the range of values reported in literature.  We 
demonstrate that the important expression
\eqref{eq:pi_est} appears to be valid within the range of uncertainty for reported
values of $\beta/C$ in figure \ref{fig:disc}.  
We also remark that the range of $\Pi$ is small with respect to values of $\beta/C$ for the Reynolds number space considered. 
Such an
apparently limited range of $\Pi$ has also been 
reported by \cite{venayagamoorthy06}. 

Furthermore, given the empirical
observation that $L_{OC}=L_O$ (figure \ref{fig:scalar_out}a) at sufficiently
high Reynolds number,
\begin{equation}
  \Pi \sim \Fr .
  \label{eq:pifr}
\end{equation}
Therefore,
the stationary Froude number 
can be thought of as a simple
consequence of the two
universal constants at high Reynolds numbers, as validated in figure
\ref{fig:disc}, which further implies $N_\ast \approx 1$,
remembering (\ref{eq:gen_hyp}).
This finally suggests that the  behaviour at lower Reynolds number may be parameterised by a
function $f_\Pi$, i.e. at unity Prandtl number
\begin{equation}
\Pi = \frac{N_\ast}{\S \sqrt{\Ri}} \approx \Fr f_\Pi(\Res).
\end{equation}

\section{Concluding remarks}
\label{sec:conclusions}
Inspired by increasing evidence that high dynamic-range stratified turbulence
exhibits 
scale-similarity as proposed by
\citet{kolmogorov41,oboukhov49,corrsin51}, we have considered the formal adaptation
of a passive-scalar mixing model \citep{beguier78} to stratified  
scalar mixing.  
By deriving the model from scaling theories, we demonstrate how model parameters have explicit functional dependence and are Reynolds number independent, at sufficiently high Reynolds number.
In order to verify our hypothesis, we have employed a model flow (i.e. S-HSST)
which equilibrates such that
the effects of decreasing dissipation scales, here parameterised by increasing shear Reynolds number $\Res$ (as defined in (\ref{eq:res})), may
be investigated independently of other flow parameters.

By direct numerical simulations of S-HSST,  we have shown that the outer length scales of turbulence and the active scalar become parametrically coupled such that the relationship between energetics and dissipation rates is predicted by the scaling theories of  Kolmogorov and Obukhov-Corrsin similarity.   
Therefore, the fundamental 
parameter associated with this relationship, $\Pi$, as defined in (\ref{eq:beg}), is the ratio (\ref{eq:pi_est}) of the
Obukhov-Corrsin constant $\beta$ and the Kolmogorov constant $C$, at sufficiently high $\Res$, and should be considered
universal to the extent that its constitutive constants are.  
As detailed in the context of various mixing models in \s 2.2, robust characterisations of $\Pi$ have strong implications on the calibration of a number of turbulence and mixing models.

This universality has profound, but still not fully-explained, implications for the turbulent flux coefficient $\Gamma$
%, and hence the `mixing efficiency' 
of such steady stratified and sheared turbulence. From the expression \eqref{eq:gammachidef1}, $\Gamma$ tending to a universal asymptotic value can now be understood as  being due to $\Pi$ tending to a constant, which follows from the scaling theory and similarity arguments presented here, that in turn rely upon classical, and relatively well-established turbulence modelling approaches. Therefore, to `understand' physically and theoretically why the empirical modeling of \cite{osborn72} and \cite{osborn80} works so well (with $\Gamma \sim 0.2$), the remaining interesting open question to address is why the ratio of energies $R_{PK}=E_p/E_k \simeq 0.08$ for  $Re_s \gtrapprox 50$, as shown in figure \ref{fig:energy}. 

The parameter $\Pi$ may be stated exactly as a function of $N_\ast,
\S , \Ri$. However, we demonstrate that it may be effectively reduced to a
function of $\Fr$ at high Reynolds number, or a function of $\Fr$ and $\Res$ at
low Reynolds numbers for the flows considered here.  The transition of the high- and low-Reynolds number regimes is observed
approximately at $\Res\approx40$, or $\Reb\approx300$ at stationary Richardson
number, for unity Prandtl numbers.   We demonstrate that large scale characteristics of the active scalar, as parameterised by $N_\ast$, exhibit significant dependence on the Reynolds number for $\Res \lessapprox 40$.  The correlation between the transition point, in Reynolds number space, of the parameter $\Pi$ and $N_\ast$ supports the underlying assumptions of the proposed model.

A stated and significant limitation of the present study is the restriction of the model to $\Fr \sim O(1)$ and $Pr=1$. An objective of future work is to perform a similar procedure to \s 2.2, except using scaling laws which account for expanded parameter regimes \citep[for instance][]{lindborg06a,batchelor59,kunze19}.

\section*{Acknowledgements}
We acknowledge the guidance of Professor J.\ J.\ Riley, the input of Dr.  F. A. V.\
de Bragan\c{c}a Alves and helpful discussions with Dr.\ R. Ristorcelli and Dr.\ J.
A. Saenz.  This work was funded by the U.S.\ Office of Naval Research via grant
N00014-15-1-2248.  High performance computing resources were provided through
the U.S.\ Department of Defense High Performance Computing Modernization
Program by the Army Engineer Research and Development Center and the Army
Research Laboratory under Frontier Project FP-CFD-FY14-007.  This manuscript is approved for public release from Los Alamos National Laboratory as LA-UR-20-28841.

\section*{Declaration of Interests}
The authors report no conflict of interest.

\bibliographystyle{jfm} 
\bibliography{bib}

\end{document}